\documentclass[a4paper,twocolumn,superscriptaddress,11pt]{quantumarticle}

\usepackage[utf8]{inputenc}
\usepackage[english]{babel}
\usepackage{graphicx,tipa}
\usepackage{csquotes}
\usepackage{amsmath}
\usepackage{amssymb}
\usepackage{xcolor}
\usepackage{color,wasysym}
\usepackage{bbold}
\usepackage[abs]{overpic}
\usepackage[T1]{fontenc}
\usepackage{hyperref}

\usepackage{cancel}

\definecolor{redred}{HTML}{D53E4F}

\definecolor{blueblue}{HTML}{1B57B6}

\newcommand{\Id}{{\mathbb 1}}

\newcommand{\trace}{\mbox{Tr}}

\newcommand{\ie}{i.e.~}
\newcommand{\eg}{e.g.~}

\newcommand{\filling}{f_M}
\newcommand{\len}{\text{L}}

\begin{document}

\title{Finite-density phase diagram of a $(1+1)$-d non-abelian lattice gauge theory
with tensor networks}

\author{Pietro Silvi}
\affiliation{Institute for complex quantum systems \& Center for Integrated Quantum Science and Technologies (IQST), Universit\"at Ulm, D-89069 Ulm, Germany}
\affiliation{Institute for Theoretical Physics, University of Innsbruck, A-6020 Innsbruck, Austria}

\author{Enrique Rico}
\affiliation{Department of Physical Chemistry, University of the Basque Country UPV/EHU, Apartado 644, E-48080 Bilbao, Spain \& IKERBASQUE, Basque Foundation for Science, Maria Diaz de Haro 3, E-48013 Bilbao, Spain}

\author{Marcello Dalmonte}
\affiliation{Institute for Theoretical Physics, University of Innsbruck, A-6020, Innsbruck, Austria \& Institute for Quantum Optics and Quantum Information, Austrian Academy of Sciences, A-6020 Innsbruck, Austria}
\affiliation{Abdus Salam International Center for Theoretical Physics, Strada Costiera 11, Trieste, Italy}

\author{Ferdinand Tschirsich}
\affiliation{Institute for complex quantum systems \& Center for Integrated Quantum Science and Technologies (IQST), Universit\"at Ulm, D-89069 Ulm, Germany}

\author{Simone Montangero}
\affiliation{Institute for complex quantum systems \& Center for Integrated Quantum Science and Technologies (IQST), Universit\"at Ulm, D-89069 Ulm, Germany}
\affiliation{Institute for Complex Quantum Systems \& Center for Integrated Quantum Science and Technologies, Universit\"at Ulm, D-89069 Ulm, Germany}

\date{\today}

\begin{abstract}
We investigate the finite-density phase diagram of a non-abelian SU(2) lattice gauge theory in (1+1)-dimensions using tensor network methods. We numerically characterise the phase diagram as a function of the matter filling and of the matter-field coupling, identifying different phases, some of them appearing only at finite densities. For weak matter-field coupling we find a meson BCS liquid phase, which is confirmed by second-order analytical perturbation theory. At unit filling and for strong coupling, the system undergoes a phase transition to a charge density wave of single-site (spin-0) mesons via spontaneous chiral symmetry breaking. At finite densities, the chiral symmetry is restored almost everywhere, and the meson BCS liquid becomes a simple liquid at strong couplings, with the exception of filling two-thirds, where a charge density wave of mesons spreading over neighbouring sites appears. Finally, we identify two tri-critical points between the chiral and the two liquid phases which are compatible with a $SU(2)_2$ Wess-Zumino-Novikov-Witten model. Here we do not perform the continuum limit but we explicitly address the global $U(1)$ charge conservation symmetry.
\end{abstract}

%\pacs{
%Pacs go here
% 05.20.-y, % Statistical Mechanics
%05.30.-d, % Quantum statistical mechanics
%52.27.Jt, % Non Netural Plasmas ?!?!?!?!?
%61.50.-f, % Crystal structure, Bulk Crystals
% 77.80.B-, % Ferroelectric phase transitions
% 64.60.an, % Finite-size systems phase transitions
%64.70.Tg, 05.30.Rt, % Quantum phase transitions
%05.10.-a. % Computational techniques for statistical physics and nonlinear dynamics
% 64.60.ae, % Renormalization-group theory in phase transition
% 02.70.-c, % Computational techniques mathematics
% 03.67.-a, % Quantum information
%}

\maketitle

\section{Introduction}

Lattice gauge theories lie at the heart of the description of nature. They provide the theoretical ground to formulate theories ranging from the Standard Model in high energy physics \cite{STDmodel1,STDmodel2} to effective condensed matter models \cite{Condmatgauge1,Condmatgauge2}, while being able to to capture superconducting and topological states of matter. Despite the tremendous success of such theories, our capability of non-perturbative analysis -- which mostly relies on numerical techniques -- has been severely limited to zero-density scenarios. Indeed, elucidating the properties of gauge theories at finite density is notoriously challenging~\cite{Philipsen:2011fv,Forcrand:2010qf} as, differently from their zero-density counterparts where Monte Carlo simulations have been tremendously successful~\cite{Montvay1994,Creutz1997,Gattringer2010}, importance sampling is invalidated due to the so called sign problem. Similarly, quantum simulators of lattice gauge theories \cite{Tagliacozzo2013,Banerjee2013,Zohar:2013qf,Stannigel:2014bf,Wiese2014,Mezzacaposu2,Zohar:2015ys}, despite being extremely promising platforms, are still in the early stage of their development \cite{EstebanChristine,LGTqinfoera}.

In this work, we investigate the zero-temperature, finite-chemical-potential phase diagram of a SU(2) lattice gauge theory in (1+1)-dimensions using Tensor Network (TN) methods \cite{MPSZero, MPSOne, MeraOne, SchollwockAGEofMPS, RomanTN}. The finite-density phase diagrams of non-Abelian lattice gauge theories have been suggested to host paradigmatic scenarios for strongly correlated quantum matter. In condensed matter settings, SU(2) gauge theories naturally emerge as a low-energy description of the Hubbard model \cite{Condmatgauge1,wenbook}, where superconductivity at finite doping represents the equivalent of meson formation at finite chemical potential. In high-energy physics, the large density regime of quantum chromodynamics (QCD) has attracted a lot of attention especially in view of possible exotic superconducting phases which may appear in regimes close to the interior of neutron stars \cite{Colorsupercond1,Colorsupercond2}.
Here we focus on (1+1)-dimensional non-Abelian models, and specifically on the
SU(2) Yang-Mills theory cast as a quantum link model \cite{horn1981,Orland1990,Chandrasekharan1997,Brower1999}, with compact gauge fields coupled to Kogut-Susskind fermions \cite{Kogut1975}.
Firstly, we show how these models support Bardeen-Cooper-Schrieffer (BCS) liquid states of matter pairs (mesons), in a wide range of Hamiltonian parameters, and that the stability of this phase
is preserved while increasing the particle density. Since there are no bare interactions between our fermionic fields, this 'color-like-superfluidity' is solely driven by the interactions mediated by the gauge fields, very much akin to suggested scenarios in SU(2) gauge theories for the Hubbard model~\cite{Condmatgauge1}. Secondly, we analyse the strong-coupling regime, where we found that chiral symmetry is spontaneously broken, but is immediately restored at finite densities, where a liquid phase with gapless single particle excitations is present. Finally, we track the behaviour of the entanglement entropy at the transition between BCS liquid and the chiral-symmetry-broken phase. Our results signal that the transition is of second order, and is compatible with a $SU(2)_2$ Wess-Zumino-Novikov-Witten model \cite{WessZumino,witten1983global,novikov1982hamiltonian}, which has been put forward as a key scenario for gauge theories with SU(2) gauge invariance \cite{tsvelikWZNW}.

%=====================================
%
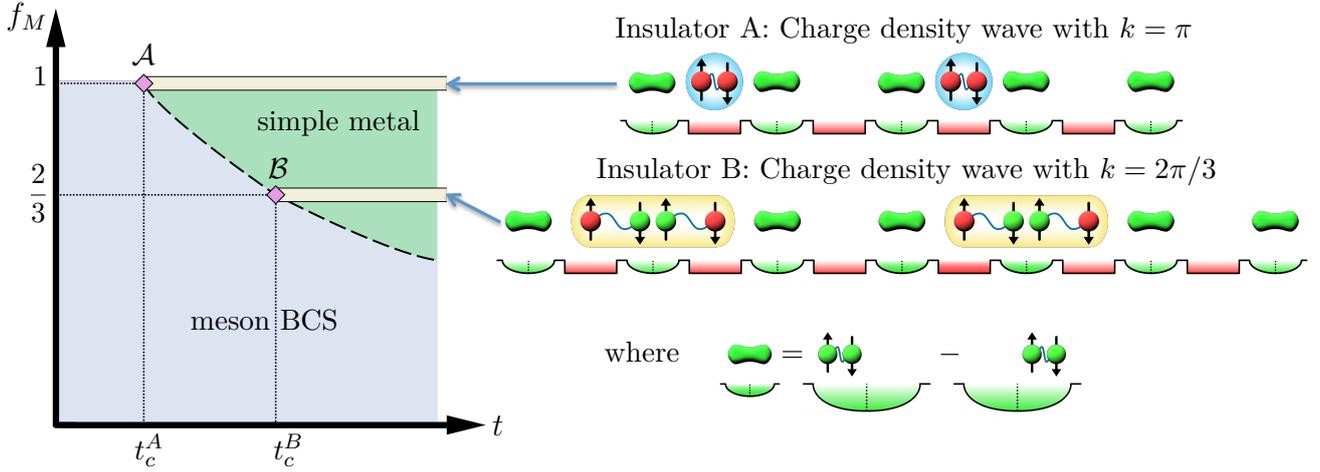
\begin{figure*}
   \begin{overpic}[width = \textwidth, unit=1pt]{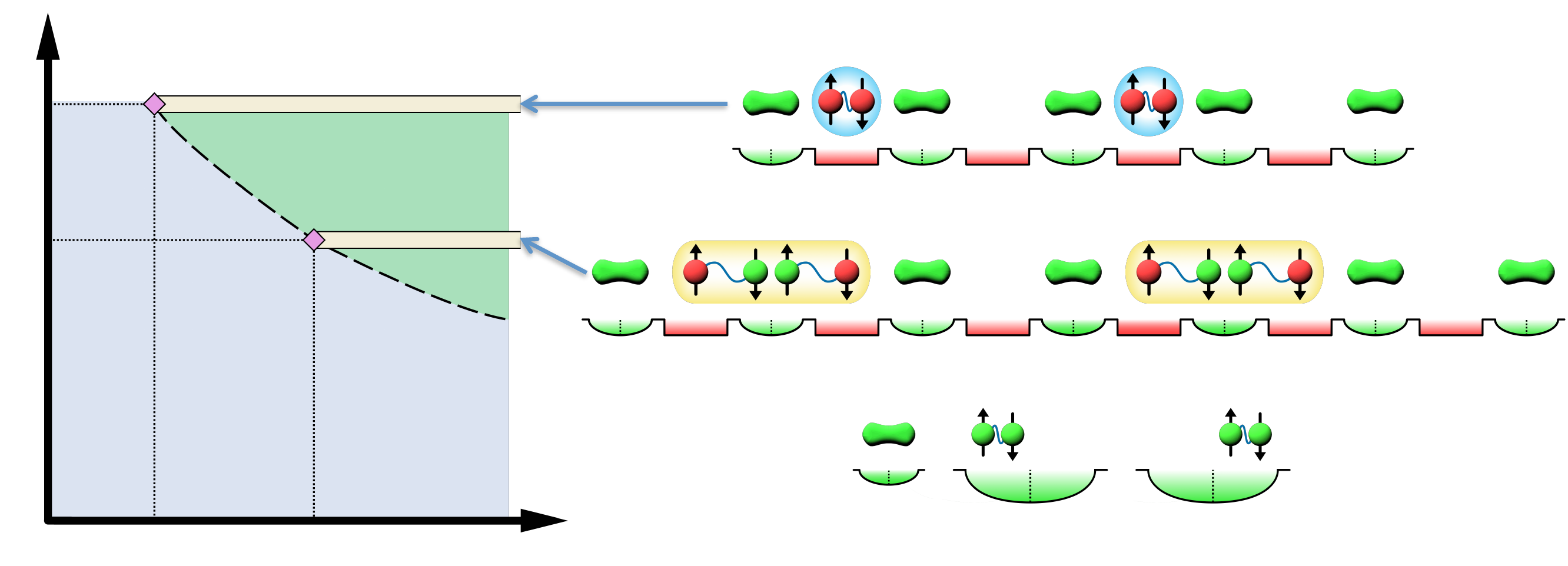}
    \put(-4, 165){\large{$\filling$}}
    \put(6, 142){$1$}
    \put(4, 98){$\displaystyle{\frac{2}{3}}$}
    \put(44, 1){$t_c^A$}
    \put(96, 1){$t_c^B$}
    \put(178, 10){\large{$t$}}
    \put(65, 50){meson BCS}
    \put(90, 125){simple metal}
    \put(43, 150){$\mathcal{A}$}
     \put(94, 107){$\mathcal{B}$}
    \put(224, 160){Insulator A: Charge density wave with $k = \pi$}
    \put(217, 107){Insulator B: Charge density wave with $k = 2\pi/3$}
    \put(220, 38){where}
    \put(286, 38){$=$}
    \put(345, 38){$-$}
   \end{overpic}
   \caption{  \label{fig:phasedia}
 Left: Phase diagram of the SU(2) lattice gauge model in quantum link formalism in the parameters space matter-field coupling and matter filling $( t, \filling)$. Two insulating phases (similar to Mott phases in Hubbard models) appears at large coupling $t$ and $\filling = 1, 2/3$. Right: cartoon states exhibiting the same local order as the insulating phases: The lattice alternates matter sites (square red wells) where matter (red circles) live, and gauge sites (round green wells, split in left and right) where rishons (green particles) live. A blue bond between two spin-1/2 particles represents a spin singlet. Insulator A alternates doubly occupied matter sites and empty matter sites, while insulator B exhibits bound states involving two adjacent matter sites (each singly occupied) and the quantum link in between, followed by an empty matter site. Rishons not entangled with matter tend to form a resonant singlet pair in the quantum link as represented in the bottom right panel.
Our simulations explored roughly a dozen of horizontal lines in the $t,\filling$ plane, with hundreds of points each and especially focusing on filling $1$ and $2/3$, plus a dozen of vertical lines, with hundreds of points each.
   }
\end{figure*}
%
%=====================================

In the last two decades, TNs, and in particular the matrix-product state methods applied here, have been widely employed to numerically tackle strongly correlated, low-dimensional many-body problems \cite{White1992,White1993,MPSZero,MPSOne,VidalTEBD,SchollwockAGEofMPS}. These wave-function based methods do not rely on importance sampling, and thus are unaffected by the sign problem, so they can be applied to zero- and finite-chemical potential regimes on equal footing. Moreover, they can easily give information about quantum correlations, such as, e.g., entanglement entropies, which have recently attracted an increasing attention in the context of gauge theories~\cite{GTEntang1,GTEntang2,GTEntang3}. As such, both in terms of regimes of applicability and observables, TNs provide a computational tool which is complementary to Monte Carlo methods. Recently, TN methods have been applied to gauge theories, with investigations ranging from the phase diagrams of Abelian theories in (1+1) and (2+1)-d, to the real time dynamics of U(1) and SU(2) models~ \cite{Byrneschwinger,Sugihara2005C,Rico2014a,psi:LGTN,Tagliacozzo2011A,Tagliacozzo2014,BanulsA,Kuhn2014,Kuhn2015,Verstraete:LGT,VerstrADD1,VerstrADD2,VerstrADD3,HanaSchwinger,Cichy2015,FuckYouAsh,meurice2013comparing,HolonomyWTF}. Here, we focus instead on the effects of finite density of fermions in a non-Abelian lattice gauge theory in the quantum link model formalism, which is very convenient for TN implementations due to its finite-dimensional Hilbert space on each link. Finally, we employ a recently developed TN structure which allows to directly work on the gauge invariant Hilbert space and to achieve the necessary numerical efficiency \cite{psi:LGTN, Tagliacozzo2014}.

From a theoretical viewpoint, our study shows how tensor methods allow to tackle questions, such as the presence of superfluid phases at finite density and the restoration of chiral symmetry in non-Abelian gauge theories, which play a fundamental role in considerably more complex theories such as QCD~\cite{QCDstories1,QDCstories2}. The analysis presented here can be extended to other gauge models, such as SU(3), more complex geometries (ladders, cylinders), while the continuum limit can be studied along the lines of recent TN results~\cite{Kuhn2014}.

\section{Results}

{\bf Quantum link formalism for SU(2)$-$} We characterise the phase diagram of a SU(2) lattice gauge theory in (1+1)-d, with spin-$\frac{1}{2}$ fermionic matter field $c^{[M]}_{j,s}$ coupled to non-abelian gauge fields $U_{j,j+1;s,s'}$ within the quantum link formulation of lattice gauge theories~\cite{Chandrasekharan1997,Brower1999,Banerjee:2013gf,Rico2014a,psi:LGTN,Mezzacaposu2}. One of the main features of quantum link formulation is that the local Hilbert space of a quantum link is finite dimensional and as such it has a finite (fermionic) representation. In this language, gauge (tensor) fields read as compound operators $U_{j,j+1;s,s'} = c^{[L]}_{j,s} c^{[R] \dagger}_{j+1,s'}$, where the 'rishon' operators $c^{[\tau] \dagger}_{j,s}$ (resp. $c^{[\tau]}_{j,s}$) create (destroy) a fermion at site $j \in \{1 .. \len\}$ and sub-lattice $\tau \in \{R,L\}$ with spin $s \in \{\uparrow, \downarrow\}$, with standard anti-commutation rules $\{ c^{[\tau] \dagger}_{j,s}, c^{[\tau']}_{j',s'} \} = \delta_{j,j'} \delta_{s,s'} \delta_{\tau,\tau'}$ and $\{ c^{[\tau]}_{j,s}, c^{[\tau']}_{j',s'} \} = 0$. They are equipped with a group of `left' and `right' transformations, generated by the $SU(2)$ 
Lie algebras
$J^{[L](\mu)}_{j,j+1} = \frac{1}{2} \sum_{s,s'= \uparrow, \downarrow} \sigma^{(\mu)}_{s,s'} c^{[L] \dagger}_{j,s} c^{[L]}_{j,s'}$
and
$J^{[R](\mu)}_{j,j+1} = \frac{1}{2} \sum_{s,s'= \uparrow, \downarrow} \sigma^{(\mu)}_{s,s'} c^{[R] \dagger}_{j+1,s} c^{[R]}_{j+1,s'}$
respectively, where the coefficients $\sigma^{(\mu)}_{s,s'}$ are the entries (row $s$, column $s'$) of the Pauli matrix $\sigma^{(\mu)}$ in direction $\mu \in \{x,y,z\}$. Within such quantum link formulation, the group of left (resp. right) transformations applies solely on the rishon mode $L$ ($R$) of the pair, and this guarantees that the two sets of generators, and thus the two groups, mutually commute $[J^{[R](\mu)}_{j,j+1},J^{[L](\mu')}_{j',j'+1}] = 0$.

Notice that, due to the fact that the fermionic representation of the link operators $U_{j,j+1;s,s'}$ and $J^{[\tau](\mu)}_{j,j+1}$ are bilinear operators, there is a local conservation of the total number of fermions within a link \cite{Chandrasekharan1997,Brower1999,Banerjee:2013gf,Rico2014a,psi:LGTN,Mezzacaposu2}. Hence, we fix the total rishon population at every link ($L_j$,$R_{j+1}$), on which the effective many-body dynamics will indeed depend. From now on, we set $\left(n^{[L]}_{j,\uparrow} + n^{[L]}_{j,\downarrow} + n^{[R]}_{j+1,\uparrow} + n^{[R]}_{j+1,\downarrow} \right) = 2$ fermions per link. Within this selection rule, every single quantum link degree of freedom has effective dimension 6, and, as discussed in Ref.~\cite{Brower1999} and reminded below, the microscopic model has a SU(2) gauge symmetry.

The local or gauge symmetry allows us to impose a SU(2)-equivalent Gauss' law at every site, which restricts the space of 'physical' states $|\Psi_{\text{phys}}\rangle$; in this case, the Gauss' law translates into the fact the total spin around every site $j$ must equal to a spin-0 (see Method's subsection).

%=====================================
%
\begin{figure*}[t]
   \begin{overpic}[width = \textwidth, unit=0.944pt]{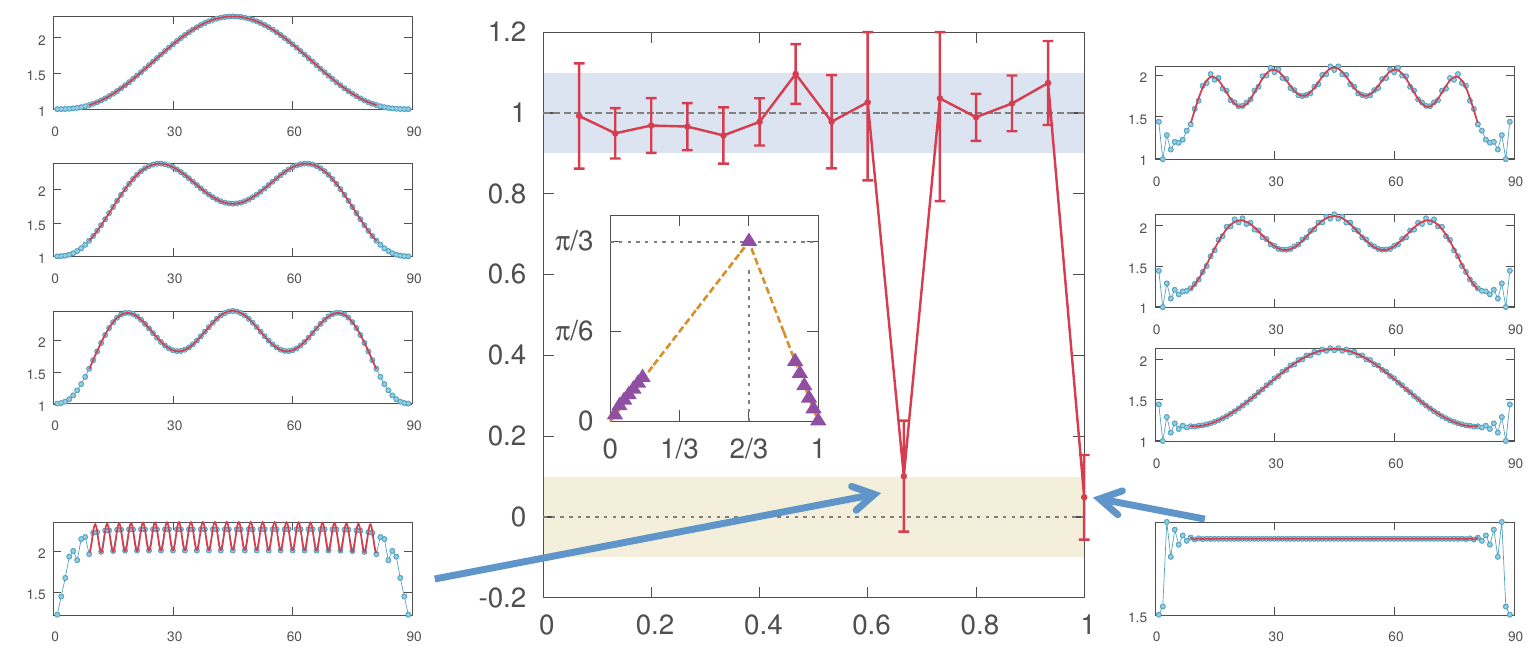}
    \put(0, 202){$S_{\ell}$}
    \put(22, 203){\scriptsize{$\filling = \frac{2}{L}$}}
    \put(120, 170){$\ell$}
    \put(0, 152){$S_{\ell}$}
    \put(120, 122){$\ell$}
    \put(36, 138){\scriptsize{$\filling = \frac{4}{L}$}}
    \put(0, 102){$S_{\ell}$}
    \put(120, 73){$\ell$}
    \put(35, 90){\scriptsize{$\filling = \frac{6}{L}$}}
    \put(80, 56){$\vdots$}
    \put(0, 32){$S_{\ell}$}
    \put(120, 2){$\ell$}
    \put(35, 19){\scriptsize{$\filling = \frac{2}{3}$}}
    \put(440, 202){$\vdots$}
    \put(366, 185){$S_{\ell}$}
    \put(480, 156){$\ell$}
    \put(400, 170){\scriptsize{$\filling = 1 - \frac{6}{L}$}}
    \put(366, 138){$S_{\ell}$}
    \put(480, 106){$\ell$}
    \put(400, 121){\scriptsize{$\filling = 1 - \frac{4}{L}$}}
    \put(366, 95){$S_{\ell}$}
    \put(480, 61){$\ell$}
    \put(412, 76){\scriptsize{$\filling = 1 - \frac{2}{L}$}}
    \put(366, 35){$S_{\ell}$}
    \put(480, 3){$\ell$}
    \put(398, 19){\scriptsize{$\filling = 1$}}
    \put(153, 173){\large{$c_0$}}
    \put(334, 3){\large{$\filling$}}
    \put(216, 188){liquid}
    \put(310, 51){insulator}
     \put(187, 121){$k_F$}
      \put(276, 70){$f_M$}
   \end{overpic}
   \caption{  \label{fig:slicer}
   (color online) Main panel: central charges $c$, fitted according to Eq.~\eqref{eq:educatedcalacardy}, for the strong-coupling theory $t = 80$, as a function of the filling $\filling$ for a chain of $\len = 90$ sites. The $c_{\text{fit}}$ values are compatible with $c = 1$ of the Luttinger liquid, thus confirming the liquid behaviour of the model, except for $\filling = 2/3$ and $\filling = 1$ where the system is noncritical (gapped) and thus insulating. 
 The error bars are a conservative estimate of the standard error from the fit, in addition to the error propagated from the tensor network simulation.
   Inset: fitted Fermi wave-vector values $k_F$, extracted via Eq.~\eqref{eq:educatedcalacardy}, as a function of the filling $\filling$ (purple dots). The orange dashed line represents the main trend, which is $\text{min}\{\frac{\pi}{2} \filling, \pi(1-\filling)\}$. Side panels: samples of bipartite entanglement $S_{\ell} = - \trace(\rho_\ell \log \rho_\ell)$ profiles (blue dots), as a function of the partition point $\ell$ along the chain, at different filling fractions $\filling$. Their corresponding
   fits according to Eq.~\eqref{eq:educatedcalacardy}, calculated after discarding 10\% of the system size at the boundaries, are shown (red curves).
   }
\end{figure*}
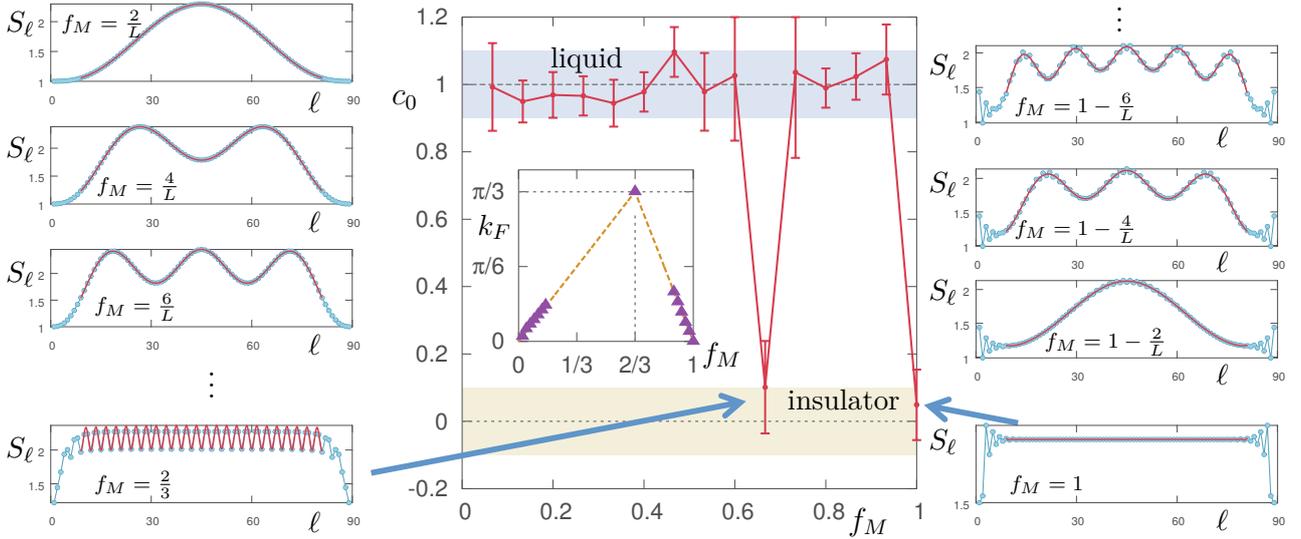
%
%=====================================

{\bf Model $-$} The Hamiltonian of model of interest encodes the microscopical dynamics of a Yang-Mills theory on a lattice, and is composed by three terms 
\begin{equation} \label{eq:ricodalmodel}
H =  H_{\text{coupl}} + H_{\text{free}} + H_{\text{break}}.
\end{equation}
 The link operators play the role of parallel transporters, and appear in the coupling term between gauge fields and matter, as \cite{Kogut1975}
\begin{equation}
 H_{\text{coupl}} = t \sum_{j = 1}^{\len-1} \sum_{s,s' = \uparrow, \downarrow} c^{[M] \dagger}_{j,s}
 U_{j,j+1;s,s'} c^{[M]}_{j+1,s'} + \mbox{h.c.},
\end{equation}
where $j \in \{1 .. \len-1\}$ and the spin $s \in \{\uparrow, \downarrow\}$. The second gauge invariant term in the Hamiltonian describes the pure-gauge field free energy
\begin{equation}
\begin{split}
 H_{\text{free}} &= \frac{g_0^2}{2} \sum_{j = 1}^{\len}  \left[ \vec{J}^{[R]}_{j-1,j} \right]^2 + \left[ \vec{J}^{[L]}_{j,j+1} \right]^2 \\
 & = 2 g_1^2
 \sum_{j = 1}^{\len} \left( 1 - n^{[L]}_{j,\uparrow} n^{[L]}_{j,\downarrow} - n^{[R]}_{j+1,\uparrow} n^{[R]}_{j+1,\downarrow} \right),
\end{split}
\end{equation}
written in terms of the fermion occupation $n^{[\tau]}_{j,s} = c^{[\tau] \dagger}_{j,s} c^{[\tau]}_{j,s}$, where $g_1 = g_0 \sqrt{3/8}$. If the quantum dynamics is ruled only by $H_{\text{free}}$ and $H_{\text{coupl}}$ there is an additional, accidental, local conservation of the number of fermions $\sum_{s = \uparrow, \downarrow} n^{[R]}_{j,s} + n^{[M]}_{j,s} + n^{[L]}_{j,s}$ around every site $j$. As a consequence, the resulting gauge symmetry of the model is $\mbox{SU(2)} \times \mbox{U(1)} = \mbox{U(2)}$~\cite{Brower1999}. In order to reduce the symmetry of the quantum link model form U(2) to SU(2) we break the additional U(1) gauge symmetry by adding the real part of the determinant of each link matrix to the Hamiltonian,
\begin{equation}
\begin{split}
H_{\text{break}} =& \frac{\epsilon}{2} \; \sum_{j=1}^{\len-1} \left( \text{det} \, U_{j,j+1} +  \text{det} \, U^{\dagger}_{j,j+1} \right) \\
=& \epsilon \sum_{j=1}^{\len-1} \left(  c^{[L]}_{j,\downarrow} c^{[L]}_{j,\uparrow} c^{[R] \dagger}_{j+1,\uparrow} c^{[R] \dagger}_{j+1,\downarrow} + \mbox{h.c.} \right) ,
\end{split}
\end{equation}
which allows the total population of every site to fluctuate. This extra component of the dynamics moves a fermion pair, which is a total spin-0 due to anti-commutation rules, from the left $L_j$ mode to the right $R_{j+1}$ model of a quantum link, or vice versa. It preserves the total spin on every site since it carries zero spin current, as well as the total population on the quantum link, keeping the symmetries previously discussed intact but the total fermion number over site $j$ and site $j+1$.
Finally, from now on we set $\hbar=1$ and $\epsilon=5$, while $g_1 \equiv 1$ sets the scale of energies.

%=====================================
%
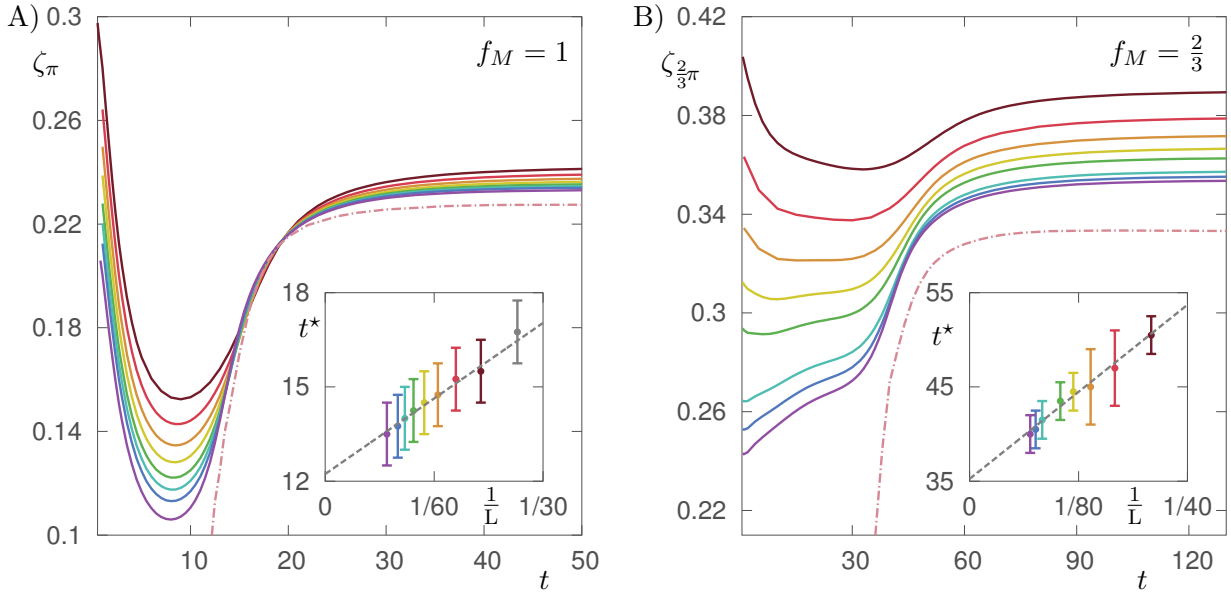
\begin{figure*}[t]
   \begin{overpic}[width = \textwidth, unit=0.9484pt]{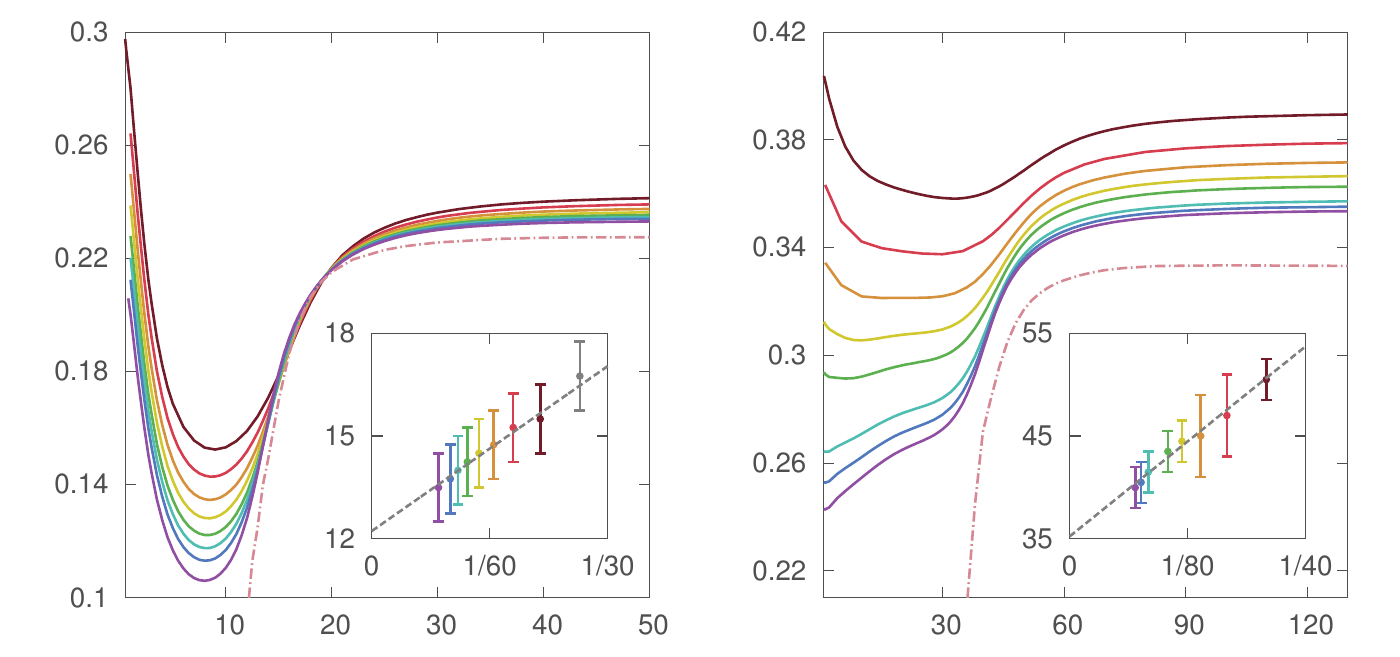}
    \put(10, 224){A)}
    \put(257, 224){B)}
    \put(20, 206){$\zeta_{\pi}$}
    \put(195, 210){$\filling = 1$}
    \put(220, 0){$t$}
    \put(197, 30){\large{$\frac{1}{\len}$}}
    \put(125, 100){$t^{\star}$}
    \put(268, 206){$\zeta_{\frac{2}{3}\!\pi}$}
    \put(445, 210){$\filling = \frac{2}{3}$}
    \put(456, 0){$t$}
    \put(451, 30){\large{$\frac{1}{\len}$}}
    \put(375, 97){$t^{\star}$}
   \end{overpic}
   \caption{  \label{fig:OPs}
   Main plots: charge density wave (CDW) order parameters $\zeta_k$ as a function of the coupling $t$. A $-$ At filling $\filling = 1$  and  $\len = 42, 50, 58, 66, 74, 82, 90, 106$ (top to bottom, brown to purple). The CDW order emerges at $k = \pi$ (chiral order). B - At filling $\filling = 2/3$ and $\len = 48,60,72,84,96,120,132,144$ (top to bottom, brown to purple) where the CDW order emerges at $k = 2 \pi / 3$. The dot-dashed curves represent the  point-by-point extrapolation to the thermodynamical limit $\len \to \infty$.
   Insets: the coupling $t^{\star}$ value at the point of steepest slope, for every curve, is plotted as a function $1/\len$ (the color code is the same as the respective main plot).
   The error bars are given by the discrete grid employed for parameter $t$.
   At the thermodynamical limit $1/\len \to 0$ we extrapolate (grey dashed line) the transition point $t^{\star} \to t_c$ with a linear fit in $1/L$, resulting in $t_c^A = 12 \pm 1$ (left) and $t_c^B = 35 \pm 2$ (right).
   }
\end{figure*}
%
%=====================================

\vspace{1em}
\textbf{Insulating Phases $-$}
Our analysis indicates the presence of two distinct gapped phases, (yellow lines in the phase diagram of Fig.~\ref{fig:phasedia}): these insulating phases possess a (different) Charge Density Wave (CDW) order, i.e.~a periodic modulation in the matter fermions populations, and thus break spontaneously translational invariance in the thermodynamical limit. The first non-critical phase (type A insulator) appears only at matter filling $\filling = 1$, where $\filling$ is the total number of matter fermions divided by the number of sites.
This phase exhibits CDW order at wave-vector $k = \pi$, \ie it has effective periodicity of 2 composite sites. The second non-critical phase (type B insulator) appears at filling $\filling = 2/3$ instead and it show an effective 3 sites periodicity (wave-vector is $k = 2 \pi /3$). These locally ordered phases arise at large matter-field coupling $t$: In the following, we report two independent methods to identify the transition point $t_c^A$ (resp. $t_c^B$).

First of all, we pinpoint these phases by investigating the bipartite entanglement quantified by the Von-Neumann entropy $S_{\ell} = \trace [\rho_\ell \log \rho_\ell]$ of the reduced density matrix $\rho_\ell = \trace_{\ell + 1 .. \len} [ |\Psi_0 \rangle \langle \Psi_0| ]$ for sites $\{1 .. \ell\}$ with $\ell < \len$, where $ |\Psi_0 \rangle$ is the ground state we find, given $g_1$, $\epsilon$, $t$ and $\filling$. The scaling of bipartite entanglement is a discriminating property between critical and non-critical theories in $1+1$ dimensions~\cite{CalabreseCardy}: While for critical systems $S_\ell \simeq S_\ell^0 = \frac{c}{6} \log ( \len \sin (\pi \ell/\len)) + c'$ diverges logarithmically with $\ell$, with $c$ being the central charge of the corresponding conformal field theory (and $c'$ a non-universal constant), for gapped systems $S_\ell$ saturates to a finite value at the thermodynamical limit. Moreover, since we are performing simulations of fermions at fixed filling $\filling$, one has to take into account Fermi oscillations as discussed thoroughly in Refs.~\cite{pasqualtweaker1,pasqualtweaker2}. In this extended framework, $S_\ell$ follows a behaviour according to
\begin{multline}
\label{eq:educatedcalacardy}
S_\ell^F \simeq S^{0}_\ell +  b_0 \cos\left( 2 k_F \left(\ell - \frac{\len}{2} \right) \right) \left( \sin\left(\frac{\pi \ell}{\len}\right) \right)^{-b_1} 
\end{multline}
where the Fermi wave-vector $k_F$ enters in the oscillatory contribution. We then use the theoretical expression~\eqref{eq:educatedcalacardy} to fit the simulated bipartite entanglement $S_\ell$ and estimate the central charge $c$, and the Fermi momentum $k_F$  of the theory. In Fig.~\ref{fig:slicer} (side panels) we plot several bipartite entanglement profiles for a strong coupling theory at various matter filling fractions $0 < \filling \leq 1$. Fermi oscillations are clearly visible, and allow us to estimate the Fermi wave-vector close to $f_M=0,1$, in particular $k_F \approx \frac{\pi}{2} \filling $ for $\filling < 2/3$, and $k_F \approx \pi (1-\filling)$ for $2/3 < \filling < 1$ as reported in the inset of Fig.~\ref{fig:slicer}. The two approximate scalings cross at $\filling = 2/3$, where, indeed, the fitted central charge $c$ has a sudden drop as shown in Fig.~\ref{fig:slicer} (central panel), and is compatible with a gapped phase (insulator B). A second drop is found at $\filling = 1$, another gapped phase (insulator A), while elsewhere the central charge is compatible to a  constant $c = 1$ signaling a Luttinger liquid phase (see next section).

The CDW order, emerging in each of the gapped phases, is signalled by the local order parameter
$\zeta_k \sim \frac{1}{\len} \sum_{j} e^{i k j} \langle n^{[M]}_j - \filling \rangle_0$
acquiring a finite nonzero value at the thermodynamical limit, for some $k$
(precisely: $k = \pi$ for $\filling = 1$, while $k = 2 \pi / 3$ for $\filling = 2/3$).
To avoid potential problems of symmetry restoration due to finite-size simulations, we extract
the CDW order parameter from the static structure factor, at wave-vector $k$, of the matter density-density correlations \cite{FisherBarber,PsiStructurefactor}
\begin{equation}
 \zeta_k \!=\! \sqrt{ \sum_{j \neq j'} \frac{e^{i k (j-j') }}{\len(\len-1)} \langle  (n^{[M]}_j - \filling)  (n^{[M]}_{j'} - \filling) \rangle_0  }
 \end{equation}
where the expectation value $\langle \cdot \rangle_0$ is calculated over $| \Psi_0 \rangle$. In Fig.~\ref{fig:OPs} we show the typical behaviour of the relevant order parameter $\zeta_k$, at filling $\filling = 1$ ($\filling = \frac{2}{3}$) in the left (right) panel as a function of the coupling parameter $t$. To identify the CDW ordered phase we compute $\zeta_k$ for different system lengths $\len$ and couplings $t$ and check when the order parameter survives at the thermodynamical limit $\lim_{\len \to \infty} \zeta_k \neq 0$.
From numerical extrapolations (see the dot-dashed curves in Fig.~\ref{fig:OPs}), it appears that the CDW order is established for large $t$ values, specifically $t > t_c^A$ (resp.~$t > t_c^B$) for $\filling = 1$
($\filling = 2/3$).
The precise identification of the transition points $t_c^A$ and $t_c^B$ between the liquid and ordered phases is performed in the next section. 

\vspace{1em}
\textbf{Transition points $-$} We can employ the CDW order parameter data to estimate the transition values $t_c^A$ and $t_c^B$, using various approaches. A first approach is the steepest slope method, which locates the transition point by taking the thermodynamical limit extrapolation $\lim_{\len \to \infty} t^{\star}$ of the parameter value $t^{\star}$ where the differential $\left. \partial \zeta_k / \partial t \right|_{t^{\star}}$ is maximal. This method is known to provide robust results as long as the critical exponent $\beta$ %(CDW primary field)
is smaller than $1$ \cite{FisherBarber}. We estimated $\beta_{A} \simeq 0.25 \pm 0.08$ and $\beta_{B} \simeq 0.13 \pm 0.05$ for the transition into insulator $A$ and $B$ respectively, via a finite-size scaling procedure (not shown) and thus we can safely apply this procedure, which results in $t_c^A = 12 \pm 1$ and $t_c^B = 35 \pm 2$  as shown in Fig.~\ref{fig:OPs} (insets). 

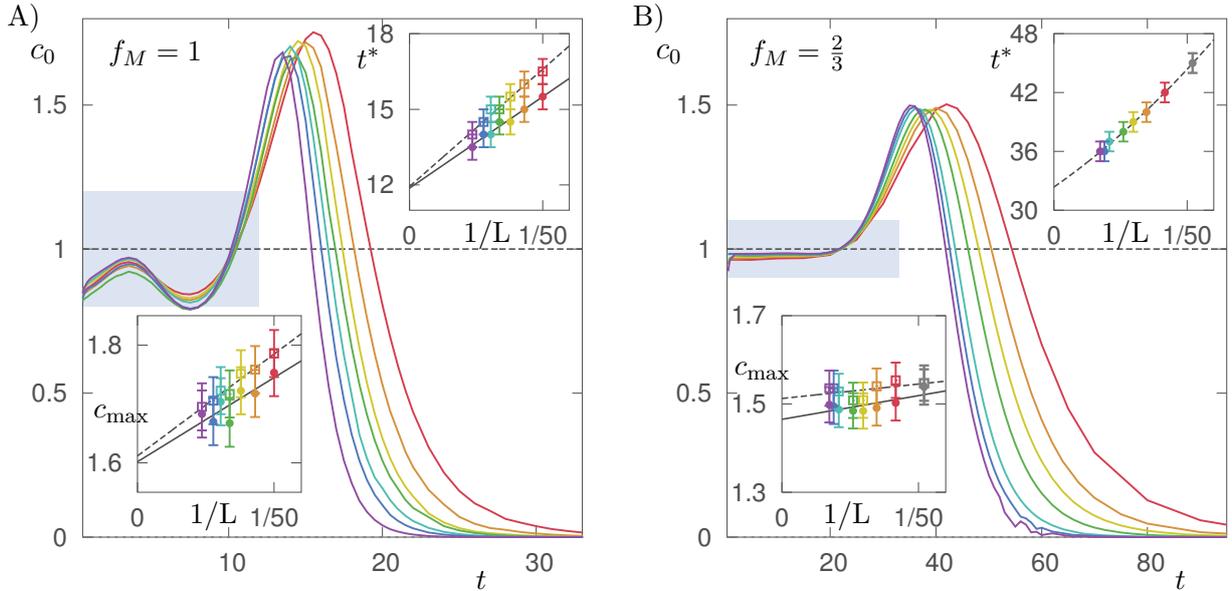
\begin{figure*}[t]
   \begin{overpic}[width = \textwidth, unit=0.9484pt]{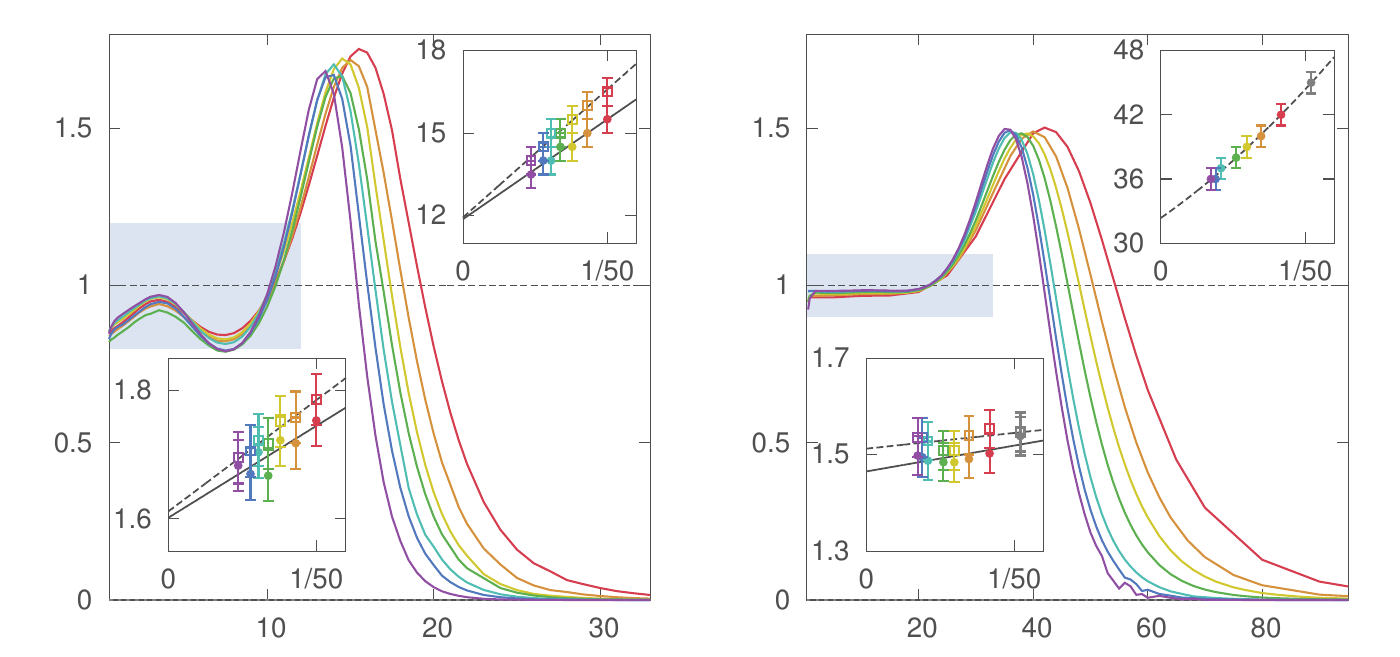}
    \put(10, 224){A)}
    \put(257, 224){B)}
    \put(19, 211){\large{$c_0$}}
    \put(50, 210){$\filling = 1$}
    \put(195, 0){$t$}
    \put(190, 138){$1/\len$}
    \put(148, 207){$t^{*}$}
    \put(82, 26){$1/\len$}
    \put(44,68){$c_{\text{max}}$}
    \put(266, 211){\large{$c_0$}}
    \put(303, 210){$\filling = \frac{2}{3}$}
    \put(471, 0){$t$}
    \put(442, 138){$1/\len$}
    \put(398, 207){$t^{*}$}
    \put(332, 26){$1/\len$}
    \put(297,85){$c_{\text{max}}$}
   \end{overpic}
   \caption{  \label{fig:charges}
   Main plots: Fitted central charges $c_0$ as a function of the coupling $t$ for various system sizes. A $-$ Filling $\filling = 1$ and $\len= 42,50,58,66,74,82,90,106$ (top to bottom, brown to purple). B $-$ Filling $\filling = \frac{2}{3}$ and $\len= 60,72,84,96,120,132,144$ (top to bottom, brown to purple). Top insets: extrapolation at the thermodynamical limit $1/\len \to 0$ of the maximal $c$ peak location $t^{*}$, which identifies the transition point: $t_c^A = 12 \pm 1$ for $\filling = 1$ and $t_c^B = 33 \pm 2$ for $\filling = 2/3$, via two independent methods (see text). Bottom insets: extrapolation at the thermodynamical limit of the maximal $c_0$ value resulting in  $c^{(A)} = 1.6 \pm 0.1$ for $\filling = 1$ and $c^{(B)} = 1.5 \pm 0.1$ for $\filling = 2/3$, extracted via two independent methods (bullet and box sets respectively, see text). In the top right inset the two data sets coincide (the color code is corresponding to the respective main plots).
   }
\end{figure*}

Another independent approach to locate the critical points is to analyse the central charge of the model as a function of the system parameters, obtained from the entanglement entropy $S_\ell$ fitted via Eq.~\eqref{eq:educatedcalacardy}, as reported in Fig.~\ref{fig:charges}.
As expected, as the thermodynamical limit is approached, a discontinuity is highlighted in the fitted $c$ values around $t_c^{A,B}$. Additionally, it appears that the transition point is described by a critical theory different from the Luttinger phases. Indeed, the spike survives at increasing system length $\len$ and leads to a two-sided discontinuity in $c$ as a function of $t$ for increasing $\ell$. Specifically, while $c \simeq 1.0 \pm 0.1$ for $t < t_c^{A,B}$ and $c \simeq 0 \pm 0.1$ for $t > t_c^{A,B}$, exactly at the transition point we extrapolate respectively $c^{(A)} = 1.6 \pm 0.1$ for the transition into insulator $A$ (at filling $\filling = 1$) and $c^{(B)} = 1.5 \pm 0.1$ for the transition into insulator $B$ (at filling $\filling = 2/3$) (lower insets in Fig.~\ref{fig:charges}). The extrapolated values $c^{(A)}$ and $c^{(B)}$ have each been obtained via two independent methods, providing compatible results: The first method is the one previously discussed, and expressed by Eq.~\eqref{eq:educatedcalacardy} (bullet points in the insets of Fig.~\ref{fig:charges}). The second method (boxes points) is based on a post-processing on the entanglement entropy profile, in order to remove the Fermi oscillations and recover a simple conformal trend $S^{0}_\ell$, whose $c$ can be easily fitted. This post-processing is somewhat technical and extensively discussed in Ref.~\cite{afflecktweaker2007}.
Our results suggest that eventual logarithmic corrections \cite{Tagliacharge1,Tagliacharge2,Tagliacharge3}, if present, do not visibly affect the quality of our fits.
(see examples shown in Fig.~\ref{fig:slicer}, obtained via the first method).
Moreover, by extrapolating the position $t^{*}$ of the peak itself at the thermodynamical limit (upper insets in Fig.~\ref{fig:charges}), we can locate the transitions points $t_c^A$ and $t_c^B$. Indeed, this procedure provides results compatible with the previous method $t_c^A = 12 \pm 1$ and $t_c^B = 33 \pm 2$. The values we obtained for the central charge are compatible with the SU(2)$_2$ WZNW model, which is an expected critical behaviour for gauge theories \cite{tsvelikWZNW}.

%=====================================
%
%
%=====================================

\vspace{1em}
\textbf{Liquid phases $-$}
Having found a wide region in the matter filling ($\filling \ne \frac{2}{3}, 1$) where the model \eqref{eq:ricodalmodel} exhibits a gapless phase with central charge $c = 1$, is already a strong signature that such phase is a band conductor (albeit an interacting one) and its long-range properties at the thermodynamical limit are captured by the Luttinger liquid paradigm \cite{bosonization,Giamarchi2003}. In this section we aim to confirm the existence of a mesonic BCS phase, where the system behaves like a Luttinger liquid of bound fermion pairs (forming a spin singlet together) rather than a liquid of single fermion quasiparticles (simple metal). This is predicted for small coupling $t$ from a perturbation theory analysis presented in the methods section. However, even for strong $t$ we also observe emergence of BCS behaviour at low fillings: In~Fig.~\ref{fig:slicer}, looking at the Fermi oscillations in the entanglement entropy, we clearly detect the shift from a scenario at $\filling \apprle 1$ where the number of oscillations is equal to the number of matter fermion holes with respect to insulator A ($k_f \approx \pi (1-\filling)$, single fermionic quasiparticle liquid), to a scenario at $\filling \apprge 0$ where the number of oscillations is equal to \emph{half} the matter fermions in the system ($k_f \approx \frac{\pi}{2} \filling$, meson BCS liquid) \cite{ManmanaTJ}.

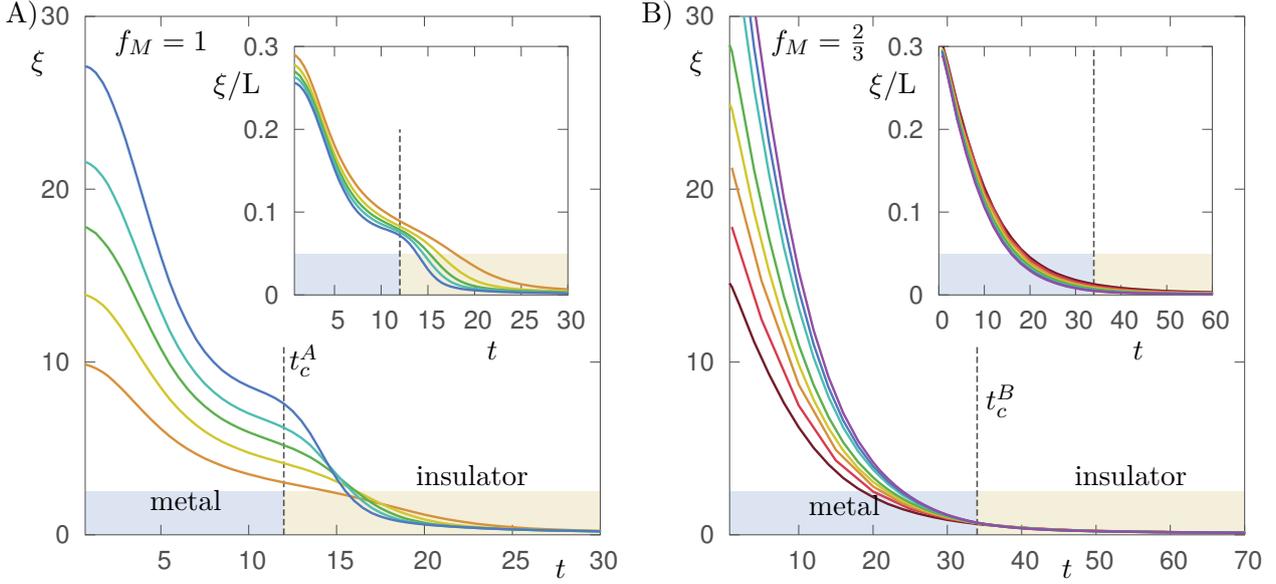
\begin{figure*}
   \begin{overpic}[width = \textwidth, unit=0.9484pt]{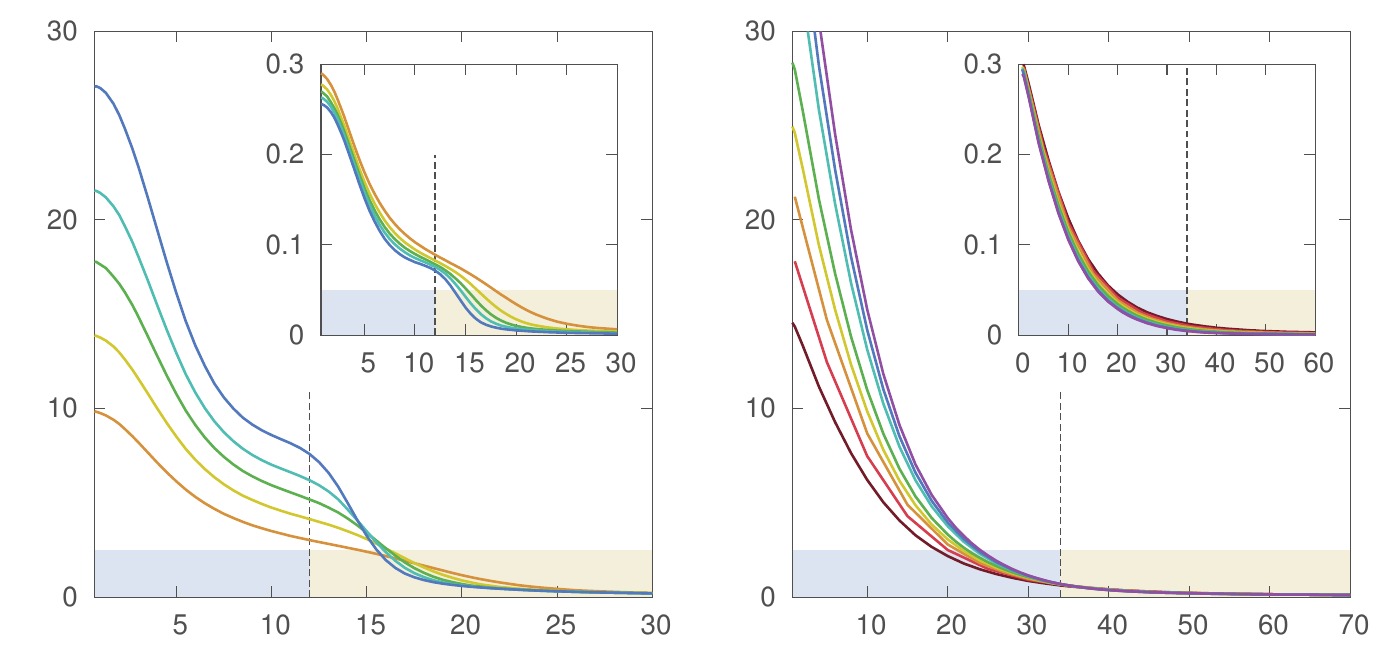}
    \put(3, 224){A)}
    \put(254, 224){B)}
    \put(46, 213){$\filling = 1$}
    \put(13, 205){$\xi$}
    \put(85, 196){$\xi/ \len$}
    \put(193, 90){$t$}
    \put(60, 30){metal}
    \put(165, 40){insulator}
    \put(115, 86){$t_c^A$}
    \put(220, 3){$t$}
    \put(305, 213){$\filling = \frac{2}{3}$}
    \put(273, 205){$\xi$}
    \put(344, 196){$\xi / \len$}
    \put(448, 90){$t$}
    \put(320, 28){metal}
    \put(425, 40){insulator}
    \put(390, 70){$t_c^B$}
    \put(420, 3){$t$}
   \end{overpic}
   \caption{  \label{fig:Corlen} 
   Main panels: Correlation lengths $\xi$ of the meson superfluid correlators, plotted as a function of the coupling $t$ and for various system lengths $\len$. A $-$ Filling $\filling = 1$  and $\len = 34,50,66,82,106$ (bottom to top, orange to indigo). B $-$ Filling $\filling = 2/3$ and $\len= 48,60,72,84,96,120,132,144$ (bottom to top, brown to purple). The phase transition points extracted from Fig.~\ref{fig:charges} and Fig.~\ref{fig:OPs} are highlighted $t_c^A \simeq 12$ for $\filling = 1$ and $t_c^B \simeq 34$ for $\filling = 2/3$. Insets: Correlation length rescaled by the total length $\len$ of the system, thus $\xi/\len$. In the gapless phases this quantity tends to a finite nonzero value at the thermodynamical limit, which signals quasi long-range order of meson superfluidity.
   }
\end{figure*}

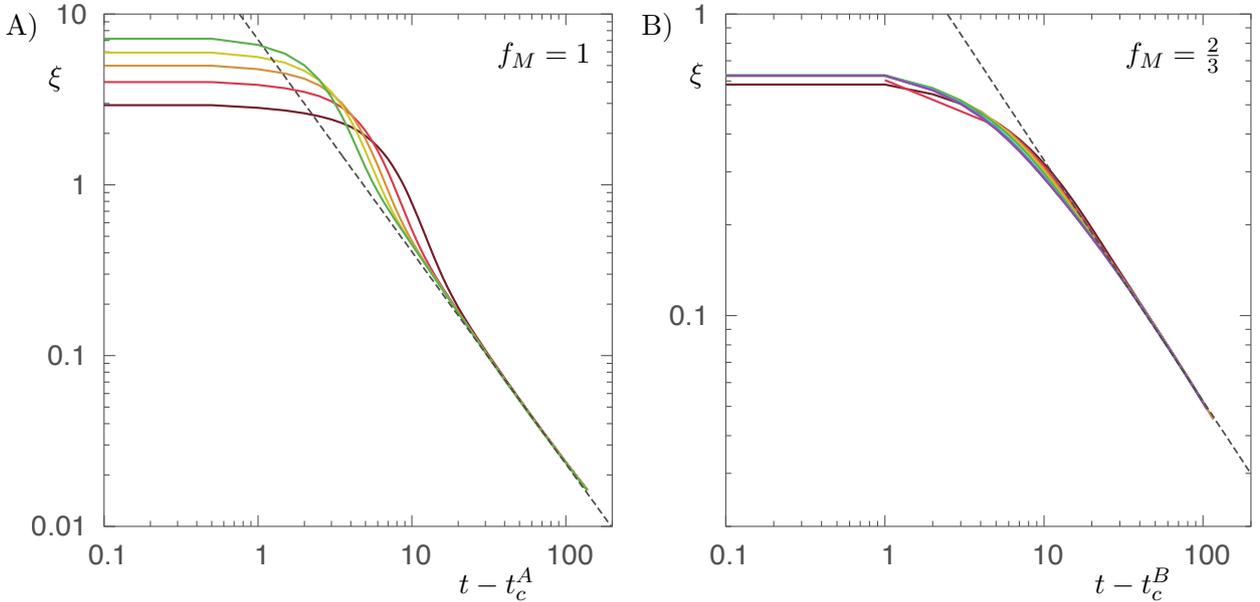
\begin{figure*}
   \begin{overpic}[width = \textwidth, unit=0.9484pt]{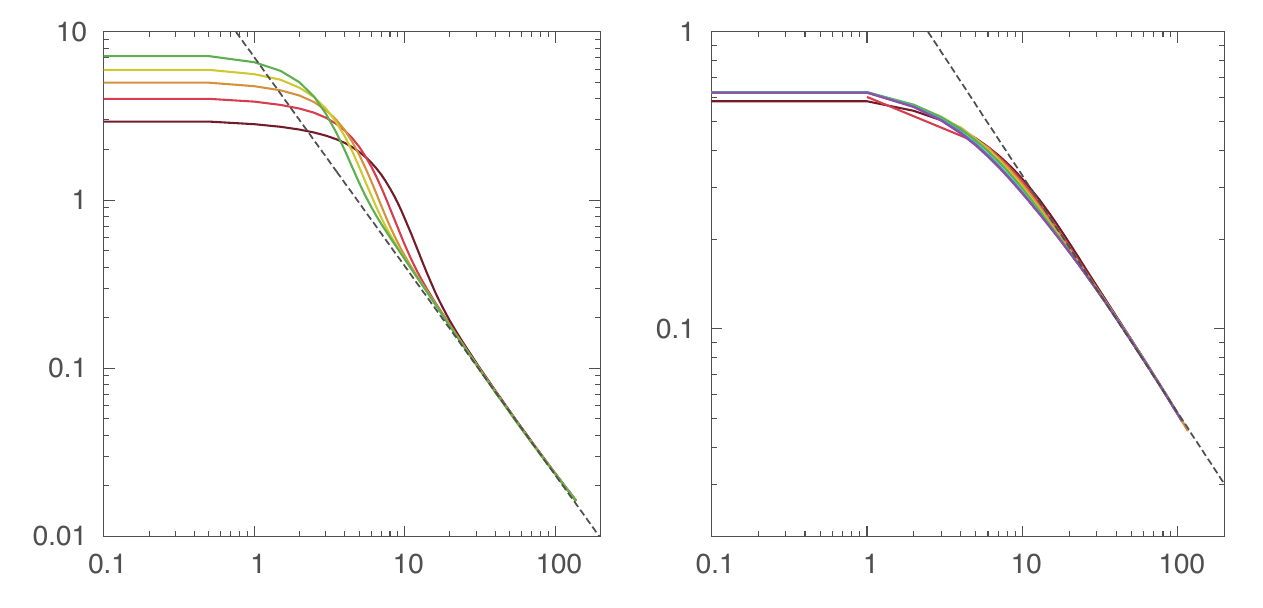}
    \put(3, 224){A)}
    \put(254, 224){B)}
    \put(197, 213){$\filling = 1$}
    \put(20, 204){$\xi$}
    \put(182, 2){$t - t_c^A$}
    \put(445, 213){$\filling = \frac{2}{3}$}
    \put(273, 205){$\xi$}
    \put(433, 2){$t - t_c^B$}
   \end{overpic}
   \caption{  \label{fig:Corlen2} 
   Main panels: Correlation lengths $\xi$ of the meson superfluid correlators, analogous to Fig.~\ref{fig:Corlen}, plotted in double logarithmic scale in the insulating phases near the respective critical point ($t_c^A$, $t_c^B$). The region where the curves follow a main trend highlights where $\xi$ is not heavily influenced by finite-size effects. In this region, $\xi$ is fitted by a power law $\xi \propto |t- t_c|^{-\nu}$ (dashed line), producing a fitted critical exponent $\nu$ corresponding to: $\nu_A \simeq 1.24$ and $\nu_B \simeq 0.79$ respectively.
   }
\end{figure*}

We confirm this picture by studying the quasi-long range order $\sigma^{-}_j = c^{[M]}_{j,\downarrow} c^{[M]}_{j,\uparrow}$ of the superfluid mesons. Since particle number conservation cannot be explicitly broken in (1+1)-d due to the Mermin-Wagner-Hohenberg theorem~\cite{Hohenberg1967,Mermin1966}, true long-range order cannot be established.
Therefore, we investigate the correlation length $\xi$ of the meson superfluid order, defined either as \cite{Kuhner2000}
\begin{equation}
 \xi = \sqrt{ \sum_{\ell \neq 0} (|\ell| - 1)^2 C_\ell / \sum_{\ell \neq 0} C_\ell },
 \end{equation}
where $C_\ell$ is the bulk average of $\langle  \sigma^{-}_j \sigma^{+}_{j+\ell} \rangle_0 = \langle c^{[M]}_{j,\downarrow} \,c^{[M]}_{j,\uparrow} \,c^{[M] \dagger}_{j+\ell,\uparrow} \,c^{[M] \dagger}_{j+\ell,\downarrow} \rangle_0$, or by fitting $C_\ell \sim a_0 \ell^{-\eta} e^{- \ell / \xi}$. We found that the two methods provide compatible results:
henceforth,  we report those obtained via the first method. 
Typical results of the correlation length are shown in Fig.~\ref{fig:Corlen}, where we specifically consider the filling values which allow for insulating phases ($\filling = \frac{2}{3}, 1$): The correlation length grows linearly with the system size $\len$ inside the whole gapless region $t < t_c^A$ ($t < t_c^B$), while it stays finite and scales as $\xi \propto (t - t_c^{A,B})^{\nu_{A,B}}$ in the insulating phase. That is, the meson BCS order survives until the insulating phase is established. An analysis of the critical exponent $\nu$ when approaching the BCS phase from the insulators is carried out in Fig.~\ref{fig:Corlen2}. 
For other filling fractions we also observed linear divergence of the correlation length, and especially at high matter filling $\filling \apprle 1$ this occurs only at small coupling $t < \bar{t}$, despite the central charge remaining constant $c = 1$. This is, again, a clear signature of the transition from the meson BCS liquid into a single fermion quasiparticle liquid, as we previously argued from the entanglement entropies.

The final phase diagram summarizing all our findings is sketched in Fig.~\ref{fig:phasedia}.

%=====================================
%
%
%=====================================

\section{Discussion}

In this manuscript we applied gauge invariant tensor networks methods to study the finite density phase diagram of a non-Abelian lattice gauge theory. This model was built to include the basic ingredients of the Yang-Mills theory for a SU(2) gauge symmetry, following a Kogut-Susskind formulation, and it has been recast in a quantum link formalism. By means of TN numerical ansatz for quantum link models, we numerically investigated the phase diagram at the thermodynamical limit. At small matter-field coupling $t$ we confirmed the existence of a meson BCS conductor, a Luttinger liquid phase of matter fermion pairs, which agrees with an analytical prediction based on second-order degenerate perturbation theory. We observed that the meson BCS may not survive at larger coupling $t$, depending on the matter filling fraction $\filling$, eventually in favor of a simple liquid phase. Moreover, chiral symmetry breaking at strong coupling leads to the presence of two insulating phases, at specific matter filling fractions $\filling = \frac{2}{3}$ and $\filling = 1$: Both these insulators exhibit charge density wave order, respectively at $k = \frac{2}{3} \pi$ and $k = \pi$. Finally, using entanglement entropies, we characterised in detail the transition points $t_c^{(A,B)}$ showing that they are compatible with a $SU(2)_2$ Wess-Zumino-Novikov-Witten critical theory. 

Our work paves the way towards non-perturbative studies of non-Abelian gauge theories in finite-density regimes where MonteCarlo simulations are biased by the sign problem. Since the local Hilbert space dimension only depends weakly on the symmetry group, it is possible to extend the present study to SU(3) models, which promise to host an even richer pairing scenario. In this context, as we have shown above, while low-dimensional models differ significantly to their higher dimensional counterparts in many-respects, they still host several common phenomena, including confinement and the restoration of chiral symmetry at finite chemical potential. The methods applied here can immediately be extended to ladder systems, and potentially to (2+1)-d lattices, providing a complementary route to recent approaches based on projected entangled paired states~\cite{Zohar:2015yq,Zohar2015}. In particular, the possibility of directly accessing entanglement properties enables the exploration of entanglement properties in gauge theories, a topic which has recently surged to attention \cite{GTEntang1,GTEntang2} and where key questions, such as the role of sub-leading corrections to the area law, promise to give tremendous insights on both critical and topological phenomena
\cite{Topoentanglement1,Topoentanglement2}.

In this study we were not interested in the continuum limit of the theory, but studied the zero temperature regime of the regularised lattice version. In this sense, our results are directly related to possible realisation in cold atom systems, where the lattice theory can be implemented \cite{EstebanChristine,Tagliacozzo2013,Zohar:2013qf}. We foresee a possible study of the continuum limit of the theory carried out via, for instance, dimensional reduction. This method has already given successful results when studying gauge theories with tensor networks, and showed that continuum limit extrapolations can be carried out even for moderately small dimension of the truncated gauge fields \cite{Verstraete:LGT,Kuhn2014,KuhnSaito,KuhnSaito2}.
In the future, we plan to investigate this aspect within the quantum link formulation we used here. 

%=====================================
%
%
%=====================================

\section{Methods}

\textbf{Gauge symmetry $-$}
By definition, gauge invariance implies that $H$ commutes with the local generators $J_j^{(\mu)}$ of $SU(2)$ gauge transformations at the site $j$ and the coordinate $\mu \in \{ x,y,z \}$, which obey $\left[ J_j^{(\alpha)} , J_{j'}^{(\beta)} \right] =  i \delta_{j,j'} \epsilon^{\alpha \beta \gamma} J_j^{(\gamma)}$. Gauge covariance dictates that, under a rotation of angle $\vec{\theta}$ at site $j$, the related matter field operators transform as
\begin{equation}
c'^{[M] \dagger}_{j,s} =e^{ -i \vec{\theta} \cdot \vec{J}_j } c^{[M] \dagger}_{j,s} e^{ i  \vec{\theta} \cdot \vec{J}_{j} }
= \sum_{s'} c^{[M] \dagger}_{j,s'} e^{- \frac{i}{2} \vec{\theta} \cdot \vec{\sigma}}_{s',s}\,,
\end{equation}
while the gauge field operators possess a left group of rotations 
\begin{equation}
 \begin{aligned}
U'_{j,j+1;s,s'} &= e^{ -i \vec{\theta} \cdot \vec{J}_j }  U_{j,j+1;s,s'}  e^{ i  \vec{\theta} \cdot \vec{J}_{j} } \\
&= \sum_{s''} \left(e^{\frac{i}{2} \vec{\theta} \cdot \vec{\sigma}}\right)_{s,s''} \; U_{j,j+1;s'',s'}
\end{aligned}
\end{equation}
and a right group or rotations 
\begin{equation}
 \begin{aligned}
U'_{j-1,j;s,s'} &= e^{ -i \vec{\theta} \cdot \vec{J}_j }  U_{j-1,j;s,s'}  e^{ i  \vec{\theta} \cdot \vec{J}_{j} } \\
&= \sum_{s''} U_{j-1,j;s,s''} \;\left(e^{-\frac{i}{2} \vec{\theta} \cdot \vec{\sigma}}\right)_{s'',s'}.
\end{aligned}
\end{equation}
This suggests that the compound gauge transformation generator $\vec{J}_j$ at site $j$ can be naturally decomposed into right-gauge, matter, and left-gauge components
\begin{equation}
 J_j^{(\mu)} = J^{[R](\mu)}_{j-1,j} + \frac{1}{2} \sum_{s,s'= \uparrow, \downarrow} \sigma^{(\mu)}_{s,s'} c^{[M] \dagger}_{j,s} c^{[M]}_{j,s} + J^{[L](\mu)}_{j,j+1}.
\end{equation}
Its commutation relations with the gauge field, which read respectively
\begin{equation}
\begin{aligned}
\left[J^{[R] (\mu)}_{j,j+1}, \!U_{j',j'+1;s,s'} \right] \!&= \!\frac{1}{2} \delta_{j,j'} \sum_{s''} U_{j,j+1;s,s''} \sigma^{(\mu)}_{s'',s'} \; ,\\
\left[J^{[L] (\mu)}_{j,j+1}, \!U_{j',j'+1;s,s'} \right] \!&= \!- \frac{1}{2} \delta_{j,j'}  \sum_{s''} \sigma^{(\mu)}_{s,s''} U_{j,j+1;s'',s'} ,
\end{aligned}
\end{equation}
are rightfully encoded in the quantum link language of the rishon modes, where we have
\begin{equation}
\begin{aligned}
\left[J^{[\tau] (\mu)}_{j,j+1}, c^{[R] \dagger}_{j',s} \right] &=  \frac{1}{2} \delta_{j+1,j'} \delta_{\tau,R}\sum_{s'} c^{[R] \dagger}_{j,s'} \;\sigma^{(\mu)}_{s',s} \; ,\\
\left[J^{[\tau] (\mu)}_{j,j+1}, c^{[L] \dagger}_{j',s} \right] &=  \frac{1}{2} \delta_{j,j'} \delta_{\tau,L}\sum_{s'} c^{[L] \dagger}_{j,s'} \;\sigma^{(\mu)}_{s',s} \;,
\end{aligned}
\end{equation}
with $\tau \in \{ L,R\}$. The Hamiltonian of Eq.~\eqref{eq:ricodalmodel} thus commutes with every $J^{(\mu)}_j$, hence it is invariant under the group of transformations $G = \exp(i \sum_{j = 1}^{\len} \vec{\theta}_j \cdot \vec{J}_j)$.

\textbf{Global symmetries $-$}
Having introduced the gauge symmetry content present in our model, let us also briefly discuss the global symmetries as well. The Hamiltonian of Eq.~\eqref{eq:ricodalmodel} clearly preserves the total amount of matter in the system
$N_{M} = \sum_{j=1}^{\len} n^{[M]}_{j,\uparrow} + n^{[M]}_{j,\downarrow}$. Moreover, since we added no chemical potential, the model is invariant under full particle-hole inversion $\Pi = (-1)^{\len} \prod_{j=1}^{\len} \prod_{\tau=R,L,M} \prod_{s = \uparrow, \downarrow} (c^{[\tau]}_{j,s} + c^{[\tau] \dagger}_{j,s})$, which is a $\mathbb{Z}_2$ group, since $\Pi^2 = \Id$. 

\textbf{Boundaries $-$}
As we simulate a one-dimensional lattice gauge system in Open Boundary Conditions (OBC), the rishon modes at the boundaries remain uncoupled, precisely $c^{[R]}_{1,s}$ and $c^{[L]}_{\len,s}$. The quantum state over these modes is left invariant by the quantum dynamics. Nevertheless, the actual space of 'physical' states is affected by the boundary conditions. Throughout our simulations, we set specific Von-Neumann boundary conditions such that there is no SU(2)-charge at the boundaries, or equivalently, no free-field energy density at the boundaries, {\it e.g.} $n^{[R]}_{1,s} |\Psi \rangle = n^{[L]}_{\len,s} |\Psi \rangle = 0$.

\textbf{Perturbation theory $-$}
When recast in the fermionic operators $c$, $c^{\dagger}$, the model defined by Eq.~\eqref{eq:ricodalmodel} reads
\begin{multline} \label{eq:ricodalmodel2}
  H = 
  g_1^2 \sum_{j=1}^{\len} ( n^{[R]}_{j,\uparrow} + n^{[R]}_{j,\downarrow} - 2 n^{[R]}_{j,\uparrow} n^{[R]}_{j,\downarrow}
  \\
  + n^{[L]}_{j,\uparrow} + n^{[L]}_{j,\downarrow} - 2 n^{[L]}_{j,\uparrow} n^{[L]}_{j,\downarrow} )   \\
  + t \sum_{j = 1}^{\text{L-1}} \sum_{s,s' = \uparrow, \downarrow} ( c^{[M] \dagger}_{j,s} c^{[L]}_{j,s} c^{[R] \dagger}_{j+1,s'} c^{[M]}_{j+1,s'}  \\+ c^{[M]}_{j,s} c^{[L] \dagger}_{j,s} c^{[R]}_{j+1,s'} c^{[M]\dagger}_{j+1,s'} ) \\
 + \epsilon \sum_{j=1}^{\len-1} c^{[L] \dagger}_{j,\uparrow} c^{[L] \dagger}_{j,\downarrow} c^{[R]}_{j+1,\downarrow} c^{[R]}_{j+1,\uparrow} + c^{[L]}_{j,\downarrow} c^{[L]}_{j,\uparrow} c^{[R]\dagger}_{j+1,\uparrow} c^{[R]\dagger}_{j+1,\downarrow}
\end{multline}
with $g_1 = g_0 \sqrt{\frac{3}{8}}$. This model is exactly solvable in the case of zero coupling $t = 0$: Indeed, when $H_{\text{coupl}}$ is absent, the remaining Hamiltonian contributions act trivially (as the identity operator $\Id$) upon the matter modes, while every quantum link $(L_j,R_{j+1})$ does not interact with any other link. Therefore, at zero temperature, every quantum link is in the state $|\text{Link}_0\rangle = |0\rangle_{j,L} |2\rangle_{j+1,R} - |2\rangle_{j,L} |0\rangle_{j+1,R}= ( c^{[L] \dagger}_{j,\uparrow} c^{[L] \dagger}_{j,\downarrow} - c^{[R] \dagger}_{j+1,\uparrow} c^{[R] \dagger}_{j+1,\downarrow}  ) | 00 \rangle$. Such a quantum link state, which is also graphically sketched in Fig.~\ref{fig:phasedia} (bottom-right panel), minimises $H_{\text{free}}$ and $H_{\text{break}}$ separately, and its energy density is $- \epsilon$. The full ground state will thus be a tensor product state where rishon modes are in $|\text{Link}_0\rangle$ state and matter models are in any arbitrary state which satisfies the Gauss' law: namely, either in $|0\rangle_{j,M}$ or in $|2\rangle_{j,M}$. This ambiguity produces a ground state degeneracy of $2^{\len}$ configurations, each of them with total ground energy $E_0 = - (\len - 1) \epsilon$. Such degenerate ground space states can be classified as spin-$\frac{1}{2}$ chain states, according to
\begin{multline}
|\Psi^{m_1 \ldots m_\len}_0\rangle = (\sigma^{+}_{1})^{m_1} \cdots (\sigma^{+}_{\len})^{m_\len} |\downarrow \downarrow \ldots \downarrow \rangle \\ 
\equiv (c^{[M]\dagger}_{1 \uparrow} c^{[M]\dagger}_{1 \downarrow})^{m_1} \ldots (c^{[M]\dagger}_{\len \uparrow} c^{[M]\dagger}_{\len \downarrow})^{m_\len} |0\ldots0\rangle_M,
\end{multline}
where $\{m_1 .. m_{\len} \}$ is any binary string and $\sigma^{-}_j = c^{[M]}_{j,\downarrow}\, c^{[M]}_{j,\uparrow}$ obey the standard Lie algebra of Pauli lowering operators: $[\sigma^{-}_j,\sigma^{-}_{j'}] = 0$, $[\sigma^{+}_j,\sigma^{-}_{j'}] = \delta_{j,j'} \sigma^{z}_j$: basically $\sigma^{+}_j = (\sigma^{-}_j)^{\dagger}$ creates a pair of matter fermions at site $j$, which we will call `meson' in analogy to high-energy physics, and possesses hardcore boson statistics. 

For small coupling regimes, $t \ll g_1^2, \epsilon$, it is possible to develop a degenerate perturbation theory analysis. First order perturbation theory leads to no contribution whatsoever. Second order degenerate perturbation theory allows us to construct an effective Hamiltonian $H_{\text{eff}}$ over the unperturbed ground space, such that $\langle \Psi_0^{\vec{m}} | H_{\text{eff}} | \Psi_0^{\vec{m}'} \rangle= \sum_{\varepsilon} (E_0 - E_{\varepsilon})^{-1} \langle \Psi_0^{\vec{m}} | H_{\text{coupl}} | \varepsilon \rangle \langle \varepsilon | H_{\text{coupl}} | \Psi_0^{\vec{m}'} \rangle$, where the sum runs over the excited states $|\varepsilon\rangle$ of the unperturbed ($t = 0$) Hamiltonian. After a lengthy but straightforward algebra, the resulting $H_{\text{eff}}$ can be rewritten in the spin-$\frac{1}{2}$ language introduced before as an antiferromagnetic Heisenberg model 
\begin{multline} \label{eq:effheisen}
H_{\text{eff}} = \frac{t^2}{2 g_1^2 + \epsilon} \sum_{j = 1}^{\len - 1} \left( \sigma_{j}^{x} \sigma_{j+1}^{x} \right.\\  \left. + \sigma_{j}^{y} \sigma_{j+1}^{y} + \sigma_{j}^{z} \sigma_{j+1}^{z} -1 \right).
\end{multline}
As the Heisenberg model is exactly solvable by Bethe ansatz, a gapless phase is realised in the limit $t \to 0$. Finally, exploiting the mapping into a Hubbard-like Model, we expect -- as confirmed by our numerical analysis -- a phase of superfluid mesons for small matter-field coupling, which can be equivalently interpreted as a BCS conductor (since mesons are matter fermion pairs).

%=====================================
%
\begin{figure}[t]
   \begin{overpic}[width = \columnwidth, unit=1pt]{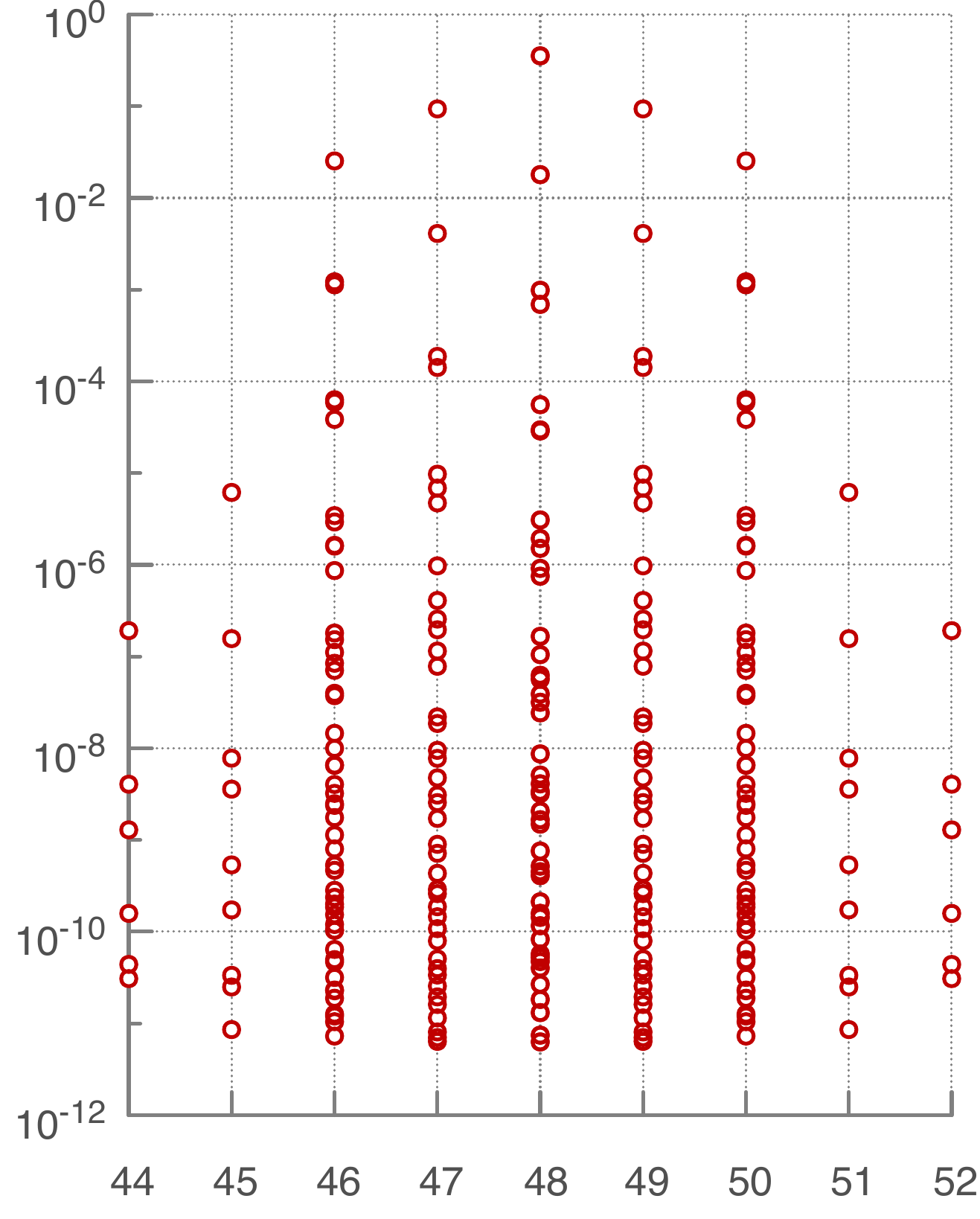}
    \put(210, -2){$q$}
    \put(-2, 266){$\lambda_{q,\alpha}$}
   \end{overpic}
   \caption{  \label{fig:entspec} 
   Entanglement spectrum $\lambda_{q,\alpha}$ of the ground state, under a partition executed at half the system size
   (i.e.~performed between sites 72 and 73 in a chain of $\len = 144$ sites). Here we are considering
   a simulation with bond link dimension $\chi=300$ in the gapless BCS phase ($t=3.5$, $\filling = 2/3$).
   The 300 values of the entanglement spectrum are resolved in the quantum number $q$ related to the
   global symmetry of matter conservation. The smallest entanglement spectrum values we
   capture with this MPS ansatz are of the order of $\sim 10^{-11}$.
   }
\end{figure}
%
%=====================================

\vspace{1em}
\textbf{Gauge invariant tensor networks $-$}
We simulate numerically the model of Eq.~\eqref{eq:ricodalmodel} in equilibrium at zero temperature,\ie in the ground state, via variational algorithm based on tensor network ansatz states for lattice gauge models introduced in Ref.~\cite{psi:LGTN}. More precisely, we adopt a Matrix Product State (MPS) ansatz \cite{MPSZero,MPSOne,MPSReps,SchollwockAGEofMPS,White1992,MPSisDMRG}, with additional embedding of both lattice gauge and global symmetries. Each simulation is performed on OBC, for finite lattice size $\len$. Numerical convergence is tested as a function of the various numerical approximation parameters (\eg the MPS bond link dimension $\chi$, see later on). The thermodynamical limit is then estimated by extrapolations to $\len \to \infty$ of the various observable quantities. In conclusion, the equilibrium properties of Hamiltonian \eqref{eq:ricodalmodel} are studied as a function of parameters $g_1$, $t$, $\epsilon$ and finally the matter filling $f_M = N_M/\len$, which we can control in a canonical fashion.

Hereafter we review and discuss the simulation details and the approximations involved. We first introduce the convention for classifying and labelling the local (canonical) basis states: we consider the three-mode compound site $j$ of the lattice (precisely, modes $R_j$, $M_j$ and $L_j$) as a single site $j$ for our simulation. Within this space, we construct the effective local basis $\{|b\rangle_j\}$, which includes all and only the $d = 14$ invariant states $|b\rangle_j$ which satisfy the Gauss' law $|J_j|^2 | b \rangle_j = 0$, where $|J_j|^2 = \sum_{\mu = x,y,z} J_j^{(\mu) 2}$. Starting from this construction of the local basis space, we build a quantum many-body (QMB) tailored wave-function ansatz based on the Matrix Product State formulation
\begin{multline} \label{eq:MPS}
|\Psi_{\text{QMB}} \rangle = \sum_{b_1 \ldots b_{\len} = 1}^{14} \sum_{\alpha_1 \ldots \alpha_{\len-1} = 1}^{\chi} \mathcal{A}^{[1] b_1}_{\alpha_1} \mathcal{A}^{[2] b_2}_{\alpha_1 \alpha_2} \mathcal{A}^{[3] b_3}_{\alpha_2 \alpha_3} \\ %\ldots \\
\ldots \mathcal{A}^{[\len] b_{\len}}_{\alpha_{\len}-1} | b_1 \ldots b_{\len} \rangle,
\end{multline}
with OBC, where the variational information is conveniently stored in rank-3 tensors $\mathcal{A}^{[j]}$. The variational complexity of ansatz \eqref{eq:MPS} is determined by $\chi$, which plays the role of a refinement parameter, and is often referred to as bond link dimension.

Finally, we encode the additional symmetries within this variational picture: the global particle conservation, which translates into the selection rule $\sum_{j} n^{[M]}_j = N_M$, and the link symmetry, whose selection rule reads $n^{[L]}_j + n^{[R]}_{j+1}$ for every $j$. This is accomplished by associating quantum numbers to the auxiliary indices $\alpha_j$ \cite{MPSsym1,MPSsym2,psi:LGTN}, thus considering $\alpha_j = (q_j, \ell_j, t_j)$, where $q_\ell$ tracks the matter particles number, $\ell_j$ carries information on the link symmetry at link $(L_j,R_{j+1})$, and $t_j$ contains all the residual information (symmetry degeneracy index). Using this formalism one obtains 
\begin{multline} \label{eq:symMPS}
\mathcal{A}^{[j] b_j}_{\alpha_{j-1}, \alpha_{j}} = \mathcal{A}^{[j] n_j^{[R]},n_j^{[M]},n_j^{[L]} }_{q_{j-1},\ell_{j-1},t_{j-1} ; q_{j},\ell_{j},t_{j} }=\\
= \underbrace{\delta_{q_{j}, q_{j-1} +n^{[M]}_j}}_{\text{global sym.}} \underbrace{ \delta_{n_j^{[R]}, 2 - \ell_{j-1}} \delta_{n_{j}^{[L]},\ell_{j}} }_{\text{link sym.}} \mathcal{R}^{[j] n_j^{[R]},n_j^{[M]},n_j^{[L]}}_{q_j,t_{j-1},t_{j+1}},
\end{multline}
where the first Kronecker delta fixes the matter conservation symmetry (once we set $q_0 = 0$ and $q_L = N_M$), and the other deltas fix the link symmetries. The residual tensors $ \mathcal{R}^{[j]}$ contains the variational parameters to be optimised.

%=====================================
%
\begin{figure}
   \begin{overpic}[width = \columnwidth, unit=1pt]{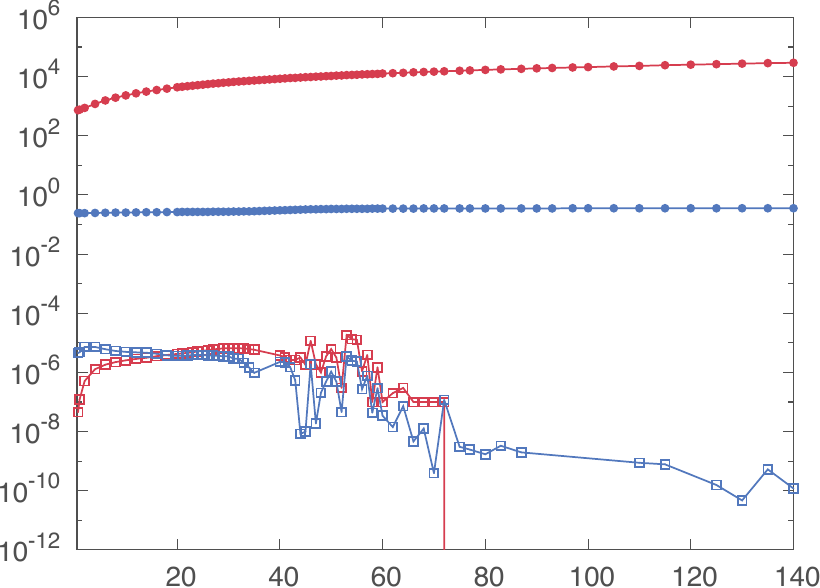}
    \put(209, -4){$t$}
    \put(70, 149){$|E_0|$}
    \put(90, 117){$\zeta_{\frac{2}{3} \pi}$}
    \put(153, 45){$\Delta \zeta_{\frac{2}{3} \pi}$}
    \put(59, 72){$\Delta E_0$}
   \end{overpic}
   \caption{  \label{fig:errors} 
   Error estimation by comparison of two simulations at different bond links, respectively $\chi = 250$ and $\chi=300$,
   carried out for several $t$ values (at filling $\filling = 2/3$).
   As relevant physical quantities here we consider the absolute value of the ground state energy $|E_0|$(red data)
   and the CDW order parameter $\zeta$ at $k=\frac{2}{3}\pi$ (blue data). The filled dots show an average of the
   two simuations, while the empty boxes show the absolute discrepancy between the simulations.
   }
\end{figure}
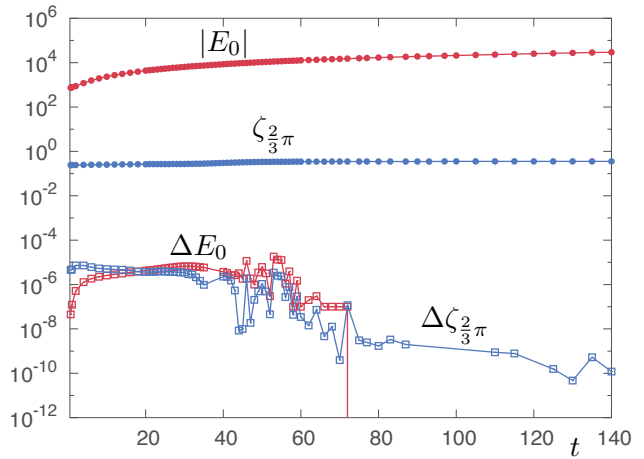
%
%=====================================

To find the ground state of Hamiltonian \eqref{eq:ricodalmodel} within the variational space given by Eqs.~\eqref{eq:MPS} and \eqref{eq:symMPS}, we adopt an annealing scheme (imaginary time evolution) implemented via the Time Evolved Block Decimation (TEBD) method~\cite{VidalTEBD,TDMRG}, evolving an initial (random) state $|\Psi_0\rangle$ in imaginary time until convergence is reached.

\textbf{Numerical resources and precision $-$}
For the system lengths that we considered $(20 \leq L \leq 150)$ we employed a MPS bond link dimension up to $\chi \sim 400$. We observed that such dimensions were sufficient to provide a robust phase analysis. Specifically, we could reach precisions of the order $\sim 10^{-11}$ in the entanglement spectra for gapless phases (as shown in Fig.~\ref{fig:entspec}), while we easily hit machine precision ($\lambda_{\min} \ll 10^{-32}$) for gapped phases.
In turn, this is expected to produce a precision of the order of $\sqrt{\lambda_{\min}} \sim 10^{-5}$ or better in the expectation values of local observables and correlators, even in the gapless phases. We confirmed this estimation by comparing simulations carried out at different bond links: Fig.~\ref{fig:errors} shows such a comparison, between $\xi = 250$ and $\xi=300$, and highlights that the errors of the quantities involved are of the order $\sim 10^{-5}$ or below.

\section{Acknowledgements}
The authors acknowledge fruitful and stimulating discussions with Peter Zoller and Uwe-Jens Wiese. SM gratefully acknowledges the support of the DFG via a Heisenberg fellowship, and
the periodic visits at IQOQI during which part of this work has been developed. 
We acknowledge support from the EU via the RYSQ, SIQS and SCALEQIT projects, the ERC Synergy Grant UQUAM, the DFG via the SFB/TRR21 and SFB FoQuS (FWF Project No. F4016-N23), and the Baden-W\"urttemberg Stiftung via Eliteprogramm for Postdocs.
We also acknowledge financial support from Spanish MINECO FIS2015-69983- P, UPV/EHU Project No. EHUA15/17, UPV/EHU UFI 11/55.

%\printbibliography
%\bibliography{su2gauge_references.bib}

\end{document}